\documentclass[twocolumn,aps,pra,groupedaddress,showpacs]{revtex4}
\usepackage{epsfig,amssymb}
\begin{document}
\title{Spin tomography}
\author{G. M. D'Ariano}
\author{L. Maccone}
\altaffiliation[also at the ]{{Massachusetts Institute of Technology,
Research Laboratory of Electronics,} MIT 36-497, Cambridge, MA 02139,
USA.}
\author{M. Paini} 
\affiliation{Quantum Optics \& Information Group, INFM Udr Pavia,\\
Dipartimento di Fisica ``A. Volta'' and I.N.F.M., Via Bassi 6, 27100
Pavia, Italy.}
\date{\today}

\begin{abstract}
We propose a tomographic reconstruction scheme for spin states.  The
experimental setup, which is a modification of the Stern--Gerlach
scheme, can be easily performed with currently available
technology. The method is generalized to multi-particle states,
analyzing the spin 1/2 case for indistinguishable particles. Some
Monte Carlo numerical simulations are given to illustrate the
technique.
\end{abstract}
\maketitle

\section{Introduction}
The main idea of tomography is to reconstruct the density matrix
$\varrho$ or, equivalently, the expectation value of any observable of
the system from repeated measurements on an ensemble of identical
states. In this paper a ``spin tomography'' for reconstructing spin
states is developed in the framework of generalized tomography,
starting from group theory {\cite{tomonew}}. There have been other
proposals to infer the spin state \cite{altri}. Our method is both
easy to carry out experimentally and for the first time allows also a
reconstruction of indistinguishable multiparticle spin 1/2 states,
which is quite general since it encloses a vast class of
experimentally accessible systems.
\par 
The best known quantum tomographic procedure is optical homodyne
tomography {\cite{tomogo}}, for the reconstruction of the density
matrix $\varrho_r$ of the radiation field from the homodyne
probability $p(x,\phi)$. It is based on the following formula
\cite{dlp}
\begin{eqnarray}\varrho_{r}=\int_0^\pi \frac{d\phi}\pi\int_{-\infty}^{+\infty} 
dx\; p(x,\phi)\;K(x-x_\phi)
\;\label{homtomog},
\end{eqnarray}
where $x_\phi$ is the quadrature operator and $K(x)$ is an appropriate
kernel function.  We will not go into details about this formula, as
we only want to stress the analogy with the spin case. In fact,
consider the spin density operator $\varrho$, which is defined on
a Hilbert space ${\cal H}_s$ of dimension $2s+1$. We will prove the
following formula
\begin{eqnarray}
\varrho=\sum_{m=-s}^{s} \int\frac{d\vec n }{4\pi}\ p(\vec n,m)
\;K_s(m-\vec s\cdot \vec n),
\;\label{tomog}
\end{eqnarray}
where the integral is performed over all directions of the versor
$\vec n$, $p(\vec n,m)$ is the probability of having outcome $m$
measuring the self-adjoint operator $\vec s\cdot \vec n$ ($\vec s$
being the spin operator), and $K_s(x)$ is a kernel function that will
be defined later.
\begin{figure}[hbt]
\begin{center}\epsfxsize=.7 \hsize\leavevmode\epsffile{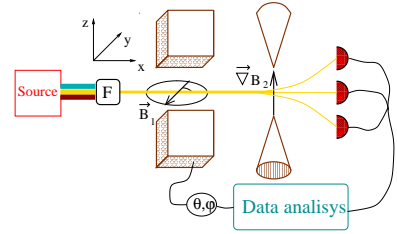}
\end{center}
\caption{Experimental apparatus for spin tomography. The Fizeau filter 
(F in the figure) selects particles with the same velocity from an
incoming beam. These are then injected into a magnetic field $\vec
B_1$, which forms an angle $\varphi$ with the $y$-axis and has
intensity proportional to $\vartheta$. For the tomographic
reconstruction the phases $\vartheta$ and $\varphi$ need to be varied
during the experiment. The remaining part of the apparatus is a
conventional Stern--Gerlach scheme (in the figure we show the case of
spin $s=1$ as an example). A computer finally correlates the
experimental results with the parameters $\vartheta$ and $\varphi$, in
order to reconstruct the density matrix, according to formula
(\ref{tomogspin}).}
\label{f:experiment}\end{figure}
\par It is possible to show that both Eq. (\ref{homtomog}) and
(\ref{tomog}) follow from a single operator identity, derived using
group theory. In fact, define Tomographic Group $G$ an unimodular
group ({\it i.e.}  left and right invariant measure are coincident)
which has a unitary irreducible square--integrable representation ${\cal
R}(g)\;(g\in G)$ on the Hilbert space $\cal H$ of the physical
system. [A square--integrable representation is such that $\int
dg\>|\langle u|{\cal R}(g)|v\rangle|^2<\infty$, where the integral is
extended to all the elements $g$ of the group $G$, $dg$ is an
invariant measure for $G$, and where the integral is not dependent on
the choice of $|u\rangle,\ |v\rangle\in{\cal H}$ (as will be shown
later)]. The operator identity we derive (\ref{homtomog}) and
(\ref{tomog}) from, and which is valid for any tomographic group $G$,
is the following
\begin{eqnarray}
\varrho =\int dg\>{\mbox Tr}[\varrho{\cal R}(g)]\; {\cal R}^\dagger(g)
\;,\label{fondfor}
\end{eqnarray}
valid for any trace class operator $\varrho$ acting on $\cal H$.  This
formula, derived from general considerations in
{\cite{tomonew,tomonew2}} is derived also in App.~\ref{s:app} using
only group theory.

The outline of the paper follows.  In Sect.~\ref{s:spspin}, the
tomography procedure to reconstruct the spin state of single particle
systems is introduced and analyzed. The experimental setup is
described and some demonstrative numerical simulations of the
procedure are studied. In Sect.~\ref{s:mpspin} the extension to the
reconstruction of multi-particle spin states is studied. For
distinguishable particles, the general reconstruction procedure is
given, while, for indistinguishable particles, the cases of two and
three spin 1/2 particles is analyzed in detail. In Sect.~\ref{s:feas}
the orders of magnitude of possible experimental setups are discussed
showing the feasibility of the proposed method. In App.~\ref{s:app}
the group tomography is derived by proving the tomographic
reconstruction formula (\ref{fondfor}) in the framework of group
theory.
\section{Single particle spin tomography}\label{s:spspin}
Starting from the general operator identity (\ref{fondfor}), we now
specify the physical system as a single spin. In this case ${\cal
H}={\mathbb C}^{\>2s+1}$, $s$ being the spin of the particle.  For
such a system, we can choose the group $SU(2)$ of $2\times 2$ unitary
matrices with unit determinant as tomographic group $G$. In fact,
$SU(2)$ can be parametrized through the ``rotation parameters'' $(\vec
n,\psi)$ ---where $\vec
n=(\cos\varphi\sin\vartheta,\sin\varphi\sin\vartheta,\cos\vartheta)$,
$\vartheta\in[0,\pi],\; \varphi\in[0,2\pi],$ and $\psi\in[0,2\pi]$---
and it induces a unitary irreducible representation on ${\mathbb
C}^{\>2s+1}$. The operators constituting this representation are given by
\begin{eqnarray} {\cal R}(\vec n,\psi)=e^{i\vec s\cdot\vec n\;\psi}
\;,\label{repres}
\end{eqnarray}
where $\vec s$ is the particle spin operator. Haar's invariant measure
\cite{haar} for $SU(2)$ is, with this parametrization and with the
normalization needed for the invariant measure (see App.~\ref{s:app}),
\begin{eqnarray} dg(\vec n,
\psi)=\frac {2s+1}{4\pi^2}\sin^2\frac\psi 2\sin\vartheta\ d\psi d\vartheta
d\varphi\;.\label{haarme}
\end{eqnarray}
As will be seen in the following, the choice of $SU(2)$ as tomographic
group $G$ is not unique.  It is easy to obtain the spin
tomography (\ref{tomog}) starting from Eq. (\ref{fondfor}), which now
rewrites as
\begin{eqnarray}
&&\varrho=\frac{(2s+1)}{4\pi^2}\int_0^{2\pi}d\psi\sin^2\frac\psi
2\int_0^\pi d\vartheta \sin\vartheta\; \nonumber\\
&&\times\int_0^{2\pi}d\varphi\;{\mbox
Tr} \left[\varrho\; e^{i\vec s\cdot\vec n\;\psi}\right]\;e^{-i\vec s\cdot\vec
n\;\psi}
\;\label{tomogspin}.
\end{eqnarray} 
Evaluating the trace over the complete set of vectors $|\vec
n,m\rangle$ (which are the eigenstates of $\vec s\cdot\vec n$,
relative to the eigenvalue $m$), we find Eq. (\ref{tomog}), by
defining \begin{eqnarray} K_s(x)\doteq
\frac{(2s+1)}\pi\int_0^{2\pi}d\psi\sin^2\frac\psi 2\; e^{i\psi x}\;,
\label{kernel}
\end{eqnarray}
and by noticing that $\langle\vec n,m|\varrho|\vec n,m\rangle=p(\vec n,m)$.

\par It should be pointed out that formula (\ref{homtomog})
for optical homodyne tomography can be proven from Eq. (\ref{fondfor})
with a very similar argument.
\par How do we use formula (\ref{tomog})? In order to measure the
matrix elements $\varrho_{il}=\langle a_i|\varrho|a_l\rangle$ for all $i,l$
($\{|a_i\rangle\}$ being a basis for ${\cal H}_s$), we only need to
calculate $\langle a_i|K_s(m-\vec s\cdot\vec n)|a_l\rangle$ and to
measure $p(\vec n,m)$.

\par The most convenient choice for the basis $\{|a_i\rangle\}$ is the
set $\{|m\rangle\}$ of eigenvectors of $s_z$ ($ m=-s,\dots,s$). Thus,
the calculation of the matrix elements of the kernel operator, by defining $\lambda_{l,m}\doteq\langle l|\vec n,m\rangle$, yields
\begin{eqnarray}
&&\langle i|K_s(m-\vec s\cdot\vec n)|l\rangle\label{matker}\\
&&=\frac{(2s+1)}\pi
\int_0^{2\pi} d\psi \sin^2\frac\psi 2 \sum_{m'=-s}^s e^{i\psi(m-m')}
\lambda_{i,m'}\lambda_{l,m'}^* \nonumber\\
&&=(2s+1)\left(\lambda_{i,m}\lambda_{l,m}^*-\frac {\lambda_{i,m+1} 
\lambda_{l,m+1}^*+\lambda_{i,m-1}\lambda_{l,m-1}^*}2\right)\nonumber
.
\end{eqnarray}
Observing that \begin{eqnarray}
|\vec n,m\rangle=e^{-i\vartheta\vec s\cdot\vec
n_\perp}|m\rangle\;\label{vecnm},
\end{eqnarray}
with $\vec n_\perp\doteq(-\sin\varphi,\cos\varphi,0)$, the evaluation
of $\lambda_{l,m}$ is given by \begin{eqnarray}
&&\lambda_{l,m}\!=\!\langle l|e^{i\vartheta(\sin\varphi
s_x-\cos\varphi s_y)} |m\rangle\!=\!\langle l|e^{-i\varphi
s_z}e^{-i\vartheta s_y}e^{i\varphi s_z}|m\rangle\nonumber\\
&&=e^{i\varphi(m-l)}\sqrt{(s+m)!(s-m)!(s+l)!(s-l)!}\nonumber\\
&&\times
\sum_\nu \frac{(-1)^\nu(\cos\frac\vartheta
2)^{2s+m-l-2\nu}(-\sin\frac\vartheta
2)^{l-m+2\nu}}{(s-l-\nu)!(s+m-\nu)!(\nu+l-m)!\nu!}\;,\label{lambda}
\end{eqnarray}
where the sum is performed over the values of $\nu$ for which the
argument of the factorials is non-negative. In the
last equality we used Wigner's formula \cite{wigner}.

\subsection{Experimental setup and state reconstruction procedure}
\par We now describe the method to measure the state $\varrho$ of an
ensemble of non-charged particles, giving the details of the
experimental apparatus, depicted in Fig. \ref{f:experiment}. The beam
of particles impinges onto a Fizeau filter, which selects one velocity
(in the $x$ direction) for the particles. This is needed in order to
ensure that each particle spends the same amount of time $t$ in the
subsequent region where a magnetic field $\vec B_1$ is present. The
field $\vec B_1$, which is parallel to the $xy$ plane, is chosen so
that $\vec B_1=B_1\vec n_\perp=B_1(-\sin\varphi,\cos\varphi,0)$. In
such way, its effect on the spin state $\varrho$ results in the
unitary transformation $U^\dag\varrho\; U$, with \begin{eqnarray}
U=\exp\left[i
\gamma B(\sin\varphi s_x-\cos\varphi s_y) t\right]\;\label{evoluz}.
\end{eqnarray}
Equation (\ref{evoluz}) follows from the Hamiltonian $H=-\vec\mu\cdot
\vec B$, with $\vec\mu\doteq\gamma\hbar\vec s$ ($\vec\mu$ being the
intrinsic magnetic moment of the particle, and $\gamma$ its
giromagnetic factor).  Successively, the particles cross a gradient of
magnetic field $\vec B_2$, whose effect is to split the beam, giving a
measure of $s_z$ for the state $U^\dag\varrho\; U$, as in a
Stern--Gerlach experiment. In this way we obtain the probability
$\langle m|U^\dag\varrho\;U|m\rangle$, which is equal to $p(\vec n,m)$
by choosing $B_1=-\vartheta/(\gamma t)$, and by using
Eq. (\ref{vecnm}). Therefore, by controlling the field $\vec B_1$, we
obtain $p(\vec n,m)$ for all $\vec n$. In fact the direction of $\vec
B_1$ selects $\varphi$, while its intensity $B_1$ selects $\vartheta$.
Now, in order to reconstruct the density matrix $\varrho$, only data
analysis is needed, {\it i.e.} the insertion of the measured $p(\vec
n,m)$ into Eq. (\ref{tomog}). One may object that an infinite number
of measures are required. However, the calculation of the integral in
(\ref{tomog}) with Monte Carlo techniques guarantees that the
reconstructed matrix elements are affected by statistical errors only,
which can be made arbitrarily small by increasing the number of
measures. In practice, a rather small number of data is required to
obtain negligible errors, as we will show by numerically simulating
the experiment.

\begin{figure}[hbt]
\begin{center}\epsfxsize=.95 \hsize\leavevmode\epsffile{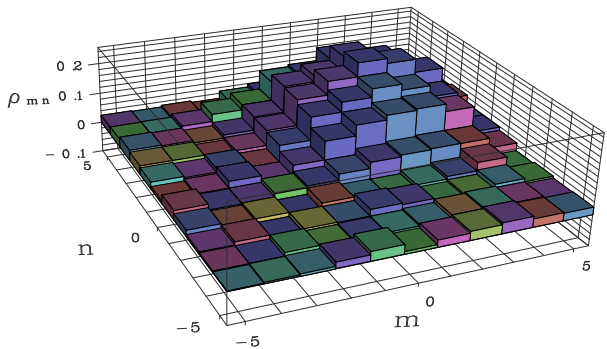}
\end{center}
\caption{Simulation of the reconstruction of the density matrix for a
coherent spin state $\varrho_{coh}$. The parameters for the state are
$\alpha=1$ and $s=5$. The simulation is performed using $3000$ spin
measurements to generate the density matrix. }
\label{f:matrcoer}\end{figure}
\begin{figure}[hbt]
\begin{center}\epsfxsize=.55 \hsize\leavevmode\epsffile{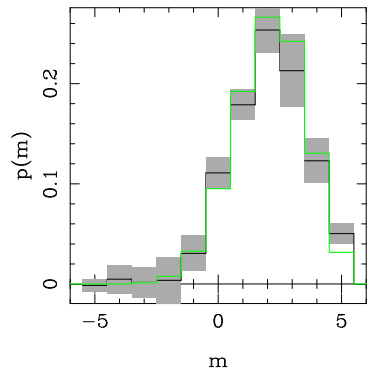}
\end{center}
\caption{Diagonal elements of the matrix given in
Fig. \ref{f:matrcoer}. The statistical error bars are obtained by
dividing the measurements into $10$ statistical blocks. The solid line
indicates the theoretical value.}
\label{f:diagcoer}\end{figure}

We first simulate the case of a coherent spin state
\cite{coher}, {\it i.e.}  \begin{eqnarray} |\alpha\rangle_s\doteq
e^{\alpha s_+-\alpha^*s_-}|-s\rangle\;,\qquad
\alpha\in{\mathbb C }\;,
\label{defcoher}
\end{eqnarray} where $s_\pm\doteq s_x\pm is_y$.
Notice the similarity with the customary optical coherent state,
defined as $|\alpha\rangle\doteq e^{\alpha
a^\dag-\alpha^*a}|0\rangle$, where $a$ is the annihilator operator for
the optical mode and $|0\rangle$ is the vacuum state.  In
Figs. \ref{f:matrcoer} and \ref{f:diagcoer} we show the reconstructed
density matrix $\varrho_{coh}=|\alpha\rangle_s {}_s\langle\alpha|$
resulting from a Monte Carlo simulated experiment.

\begin{figure}[hbt]
\begin{center}\epsfxsize=.95 \hsize\leavevmode\epsffile{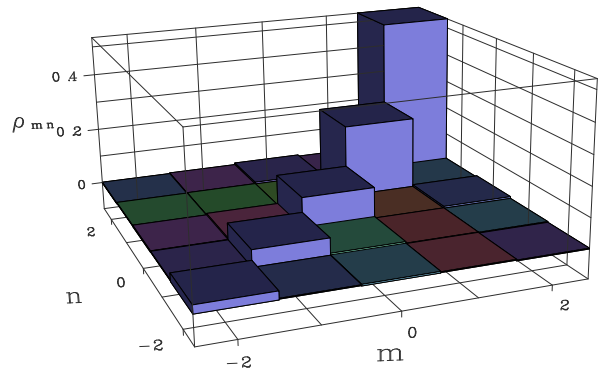}
\end{center}
\caption{Density matrix for a thermal spin state $\varrho_{th}$. 
Here $\epsilon=.75$ and $s=2$. A number of $60000$ simulated
measurements have been used in the reconstruction. }
\label{f:matrterm}\end{figure}

\begin{figure}[hbt]
\begin{center}\epsfxsize=.55 \hsize\leavevmode\epsffile{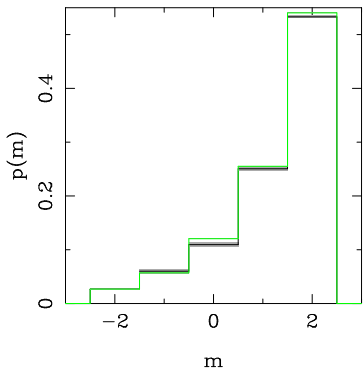}
\end{center}
\caption{Main diagonal of the matrix given in
Fig. \ref{f:matrterm}. The error bars, which in this case are
practically negligible, are obtained by dividing the measurements into
$10$ statistical blocks. The solid line indicates the theoretical
value.}
\label{f:diagterm}\end{figure}

As an additional example, in Figs.~\ref{f:matrterm} and
{\ref{f:diagterm}} we give the simulated reconstruction of a thermal
spin state, which is the mixture defined by \begin{eqnarray}
\varrho_{th}\doteq \frac{e^{-\epsilon s_z}}{{\mbox Tr}[e^{-\epsilon
s_z}]}\;,\qquad \epsilon\in{\mathbb R}\;.\label{deftherm}
\end{eqnarray}
The state $\varrho_{th}$ describes a gas of non interacting spins in
thermal equilibrium with a reservoir at a temperature $T$ and in the
presence of a magnetic field $B_z$ parallel to the $z$-axis, {\it
i.e.}  $\epsilon=-\gamma\hbar B_z/(K_BT)$, $K_B$ being the Boltzmann
constant.

\subsection{Discrete spin tomography}
Up to now $SU(2)$ has been used as tomographic group $G$ for the
reconstruction of the spin density matrix. This choice for $G$ is not
unique. For example, in the case of spin $s=\frac 12$, it is possible
to use also the group defined as ${\cal G}\doteq\{i\vec\sigma,
-i\vec\sigma, I,-I\}$, where $\vec\sigma$ is the vector of Pauli
matrices $\vec\sigma\doteq(\sigma_x,\sigma_y,\sigma_z)$. The following
irreducible unitary representation on ${\mathbb C}^{\>2}$ exists
\begin{eqnarray}
&&{\cal R}(i\sigma_\alpha)={\cal R}(-i\sigma_\alpha)=\sigma_\alpha,\
\alpha=x,\;y,\;z\nonumber\\&&
{\cal R}(I)={\cal R}(-I)=I
\;\label{rappresum}.
\end{eqnarray}
Using this representation, from the tomographic reconstruction formula
(\ref{fondfor}) we obtain
\begin{eqnarray}
\varrho=\sum_{m=-1/2}^{1/2}\sum_{\alpha=x,y,z} p(\vec n_\alpha,m)
m \sigma_\alpha+\frac 12
\;\label{rhoum}.
\end{eqnarray}
Notice that, by using Eq. (\ref{rhoum}) it is sufficient to measure the
spin in only three directions. 
\par
\begin{figure}[hbt]
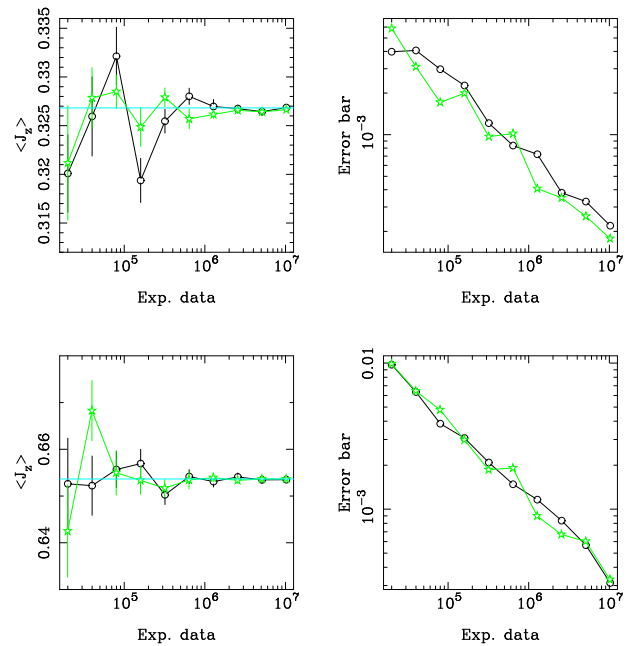

\begin{center}
\epsfxsize=.45 \hsize\leavevmode\epsffile{contvsdisc1.ps}
\hspace{.2cm}
\epsfxsize=.45 \hsize\leavevmode\epsffile{contvsdisc2.ps}
\end{center}
\begin{center}
\epsfxsize=.45 \hsize\leavevmode\epsffile{contvsdisc3.ps}
\hspace{.2cm}
\epsfxsize=.45 \hsize\leavevmode\epsffile{contvsdisc4.ps}
\end{center}
\caption{Monte Carlo comparison between continuous and discrete
tomography. Continuous tomography uses $SU(2)$ as tomographic group
and is based on Eq. (\ref{tomogspin}), while discrete tomography uses
$SU(2)$ finite subgroups and is based on the reconstruction procedures
given in Eq. (\ref{rhoum}) for $s=\frac 12$ and Eq. (\ref{rhos1}) for
$s=1$. Left: Convergence of the mean value of $\langle s_z\rangle$ for
a coherent $\alpha=2$ spin state for increasing number of experimental
data (the theoretical value is given by the horizontal lines).  The
circles refer to continuous, the stars to discrete tomography. The
upper graph is for spin $s=\frac 12$, the lower is for $s=1$. Right:
Plots of the statistical error bars of the graphs on the left {\it vs}
experimental data. The error bars are obtained by dividing the
experimental data into $20$ statistical blocks. Notice that the two
tomographic procedures are essentially equivalent: they converge in
the same way at the same result.}
\label{f:discr}\end{figure}

\noindent
Analogously, for spin $s=1$ it is possible to find a finite group in
alternative to $SU(2)$. In fact, consider the 12 element tetrahedric
group composed of the $\pm\frac 23\pi$ rotations around the versors
$\{\vec n_1=\frac 1{\sqrt{3}}(1,1,1),\>\vec n_2=\frac
1{\sqrt{3}}(1,-1,-1),\>\vec n_3=\frac 1{\sqrt{3}}(-1,1,-1),\>\vec
n_4=\frac 1{\sqrt{3}}(-1,-1,1)\}$, of the $\pi$ rotations around
$\{\vec n_5=(1,0,0),\>\vec n_6=(0,1,0),\>\vec n_7=(0,0,1)\}$ and of
the identity. It induces a unitary irreducible representation on the
space ${\mathbb C}^{\>3}$, given by the $3\times 3$ rotation
matrices. Hence, Eq. (\ref{tomog}) now becomes
\begin{eqnarray}
\varrho=\frac 14\sum_{m=-1}^1\sum_{j=1}^7p(\vec n_j,m){\cal K}_j(m-\vec
s\cdot\vec n_j)+\frac 14I
\;\label{rhos1},
\end{eqnarray}
with \begin{eqnarray}
{\cal K}_j(x)=\left\{\matrix{&2\cos(\frac 23\pi x)& \ j=1,\cdots,4\cr
&e^{-i\pi x}&\ j=5,6,7}\right.\;\label{kernels1}.
\end{eqnarray}
Notice that this procedure does not make use of a minimal set of
measurements, since 14 experimental parameters must be determined in
(\ref{rhos1}), whereas there are only 8 independent real parameters in
the $3\times 3$ density matrix. On the contrary, the case of spin
$s=\frac 12$ outlined previously does use the minimal set of
measurements for such a system. In Fig. \ref{f:discr} a comparison
between the two spin tomography procedures given by
Eqs. (\ref{tomogspin}) and (\ref{rhos1}) is shown through a Monte
Carlo simulation. Notice that there is no significant difference in
the results, showing that there is no substantial need for a procedure
which involves a minimal set of measurements.
\par\noindent
For spins $s>1$ an analogous procedure holds: one needs to find a
finite group such that it induces an irreducible unitary
representation on ${\cal H}={\mathbb C}^{\>2s+1}$.

\section{Many particle spin tomography}\label{s:mpspin}
The mathematical extension of the method to the case of a system composed
of many spins is trivial, yet, it predicts the necessity of performing
measurements on single components and this may not always be possible
when the system is composed of indistinguishable particles. 
For this reason, we need to develop the theory more.
\subsection{Distinguishable spins.}
As tomographic group for a system of $N$ spins we can simply use
$SU(2)^{\otimes N}$.  Up to equivalences, its irreducible
representations are given by the direct product of $N$ operators
(\ref{repres}) and the invariant measure is the product of $N$
measures (\ref{haarme}). As a consequence of the tomography
reconstruction formula (\ref{fondfor}) applied to $SU(2)^{\otimes N}$,
we attain readily the following generalization of
Eq. (\ref{tomogspin})
\begin{eqnarray}
&&\varrho=\prod_{k=1}^N 
\frac{(2s_k+1)}{4\pi^2}\int_0^{2\pi}d\psi_k\sin^2\frac{\psi_k}
2\int_0^\pi d\vartheta_k \sin\vartheta_k\; \nonumber\\
&&\times\int_0^{2\pi}d\varphi_k\;{\mbox Tr} \left[\varrho\; e^{i\vec
s_k\cdot\vec n_k\;\psi_k}\right]\;e^{-i\vec s_k\cdot\vec n_k\;\psi_k}
\;\label{tomogmanyspin},
\end{eqnarray}
where $k$ is the particle index. The trace term in
(\ref{tomogmanyspin}) gives rise to the probability $p(\vec
n_1,m_1;\cdots;\vec n_N,m_M)$ of obtaining $m_k$ as result for the
measurement of the $k$th spin $\vec s_k$ in the direction $\vec
n_k$. This information is accessible only in the case of fully
distinguishable spins.
\par
\begin{figure}[hbt]
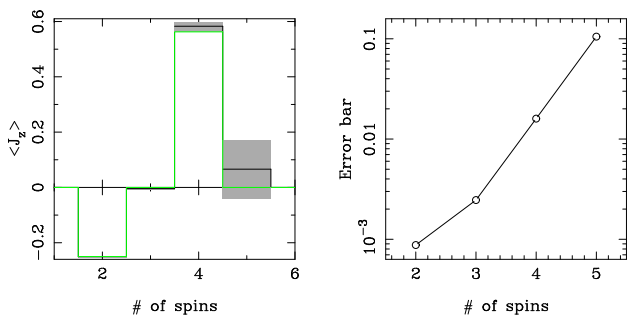

\begin{center}
\epsfxsize=.45 \hsize\leavevmode\epsffile{many.ps}
\hspace{.3cm}
\epsfxsize=.45 \hsize\leavevmode\epsffile{manyerrs.ps}
\end{center}
\caption{Left: Plot of $\langle S_z\rangle$ for different number of spins in 
a completely symmetrical state. A total of $10^6$ measurements for
each mean value was performed in this simulation. Right: Semilog plot
of the error bars {\it vs} the number of spins. Notice the exponential 
increase in the statistical errors.}
\label{f:many}\end{figure}

 In Fig. \ref{f:many} a simulated tomographic reconstruction of
the value of $\langle S_z\rangle$ ($S_z$ being the total spin
component in the $z$-direction) is given for different multiparticle
spin states. Notice how the number of the necessary experimental data
increases exponentially with the number of spins, since the
statistical error is exponential in the number of particles.

\subsection{Indistinguishable spin $1/2$ particles.}
Suppose we were given a system of $N$ particles with the same spin.
Such particles may be treated as identical by introducing a new
dynamical variable, as in the case of the isospin. The spin density
matrix (which is the partial trace over the orbital degrees of freedom
of the global density matrix) is completely symmetrical, {\it i.e.}
\begin{eqnarray}
P\varrho P^{-1}=\varrho
\;\label{symm},
\end{eqnarray}
for any particle permutation $P$, because of the complete symmetry of
the global density matrix. \par

It is also possible to see that the spin density matrix is block
diagonal in the representation of vectors of definite symmetry, with
the subspace corresponding to each block spanned by vectors belonging
to the same symmetry. In fact, given $|\phi\rangle$ and $|\psi\rangle$
vectors of different symmetry type
\cite{hamer}, then $\langle\phi|\psi\rangle=0$. Hence, for any
operator $\varrho$, satisfying (\ref{symm}), one has $\langle
\phi|\varrho|\psi\rangle=0$, as $\varrho|\psi\rangle$ belongs to
the same symmetry type as $|\psi\rangle$. \par

Since the square of the total spin $S^2$ and its $z$ component $S_z$
both commute with all permutation operators $P$, the common
eigenvectors of $S^2$ and $S_z$ may be taken as a base for each of the
diagonal blocks of the spin matrix. Let us now restrict our attention
to $s=1/2$ spin particles. In this case, to each symmetry type there
corresponds only one value of $S$, where $S(S+1)$ is the eigenvalue
of $S^2$. In fact, given $[\lambda_1\lambda_2]$ the partition of $N$
which defines the class of permutations $P$ that indicate a symmetry
type, we find $S=\frac 12(\lambda_1-\lambda_2)$ \cite{messiah}. \par

Let ${\cal H}_{S,M}$ be the space of vectors with assigned $S$ and $M$
($M$ being the eigenvalue of $S_z$). The spin density matrix
restricted to ${\cal H}_{S,M}$, which is given by $\varrho_{S,M}$, is
again completely symmetrical, hence $[P,\varrho_{S,M}]=0$. Moreover,
${\cal H}_{S,M}$ is associated with an irreducible representation of
the permutations group \cite{messiah}. By using Schur's lemma, we can
thus conclude that $\varrho_{S,M}\propto I$, $I$ being the identity in
${\cal H}_{S,M}$. In ${\cal H}_{S,M}$ there may be vectors of
different symmetry type $i$, yet $\langle i,S,M| \varrho_{S,M}
|i,S,M\rangle$ does not depend on the index $i$, so that the
probability for the measurement of $S^2$ and $S_z$ does not depend on
the symmetry type. The same conclusion holds for the measurement of
$S^2$ and $\vec S\cdot m$ for any versor $\vec m$. Hence, from the
arbitrariness of $\vec m$, we conclude that blocks with the same $S$
(and different symmetry type) are coincident.  \par

In conclusion, we have proved that in the $\{S^2\}$ representation
$\varrho$ is block diagonal, that each block corresponds to a value of
$S$ and that blocks with the same $S$ are equal. Remarkably, applying
Eq. (\ref{tomog}) to each block, we can reconstruct $\varrho$
measuring only the global quantities $S^2$ and $\vec S\cdot\vec
n$. Some examples will clarify both the theory and the needed
experimental setup.

\par
\begin{figure}[hbt]
\vskip -1truecm
\begin{center}
\epsfxsize=1 \hsize\leavevmode\epsffile{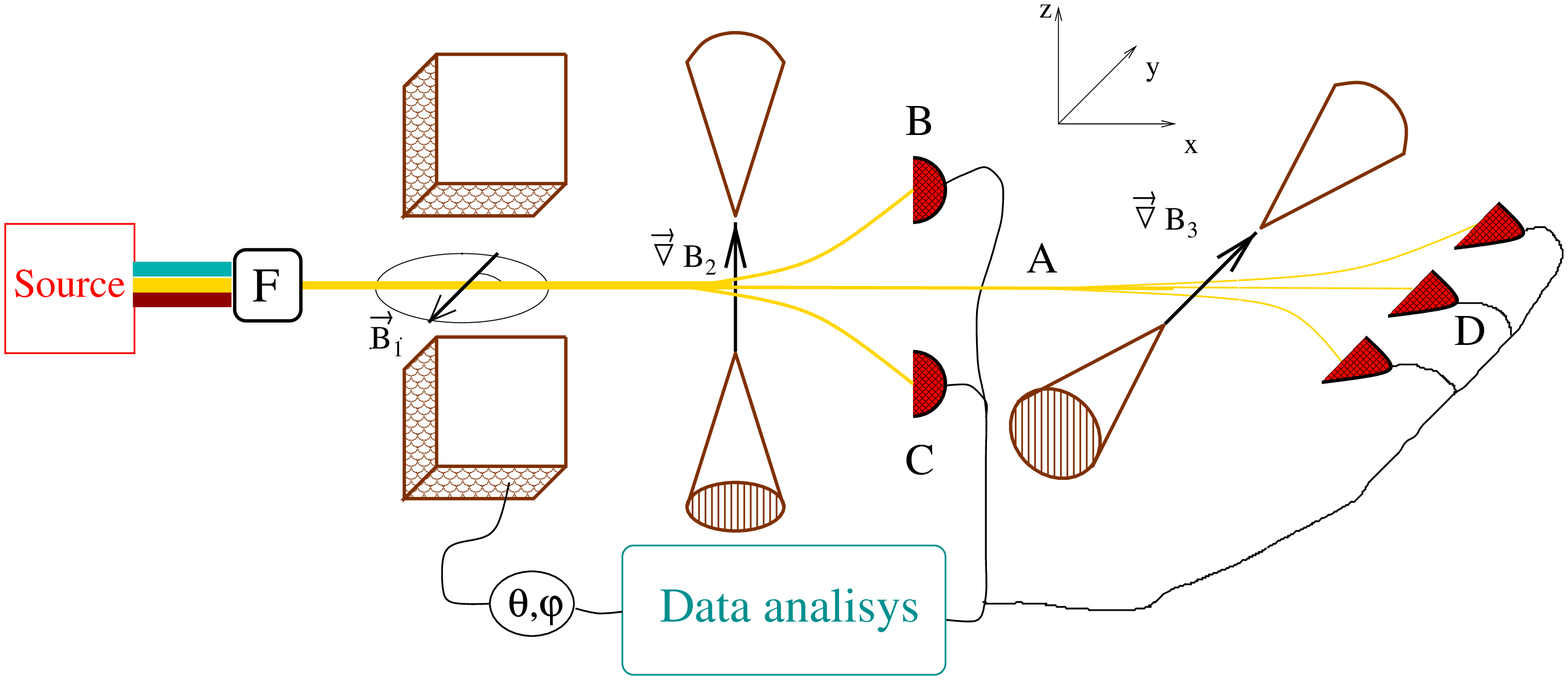}
\end{center}
\caption{Experimental apparatus for the tomography of systems composed 
of two spins $\frac 12$.}
\label{f:experiment2part}\end{figure}

 In the case of two spins $1/2$, the spin density matrix will
be of the form
\begin{eqnarray}
\varrho=
\left( \begin{array}{cccc}
\sigma_{11} & \sigma_{12} & \sigma_{13} & 0 \\
\sigma_{21} & \sigma_{22} & \sigma_{23} & 0 \\
\sigma_{31} & \sigma_{32} & \sigma_{33} & 0 \\
0 & 0 & 0 & \alpha 
\end{array} \right)\doteq \sigma\oplus\alpha\;,\label{matrix}
\end{eqnarray}
where the $\sigma$ block corresponds to the subspace spanned by the
eigenstates of $S=1$ (which are symmetrical with respect to particles
permutations), while the $\alpha$ block to the subspace spanned by the
only eigenvector of $S=0$ (anti--symmetrical with respect to
permutations). Applying (\ref{tomog}) to each block one finds
\begin{eqnarray}
\varrho&=&\!\int\frac{{{{\mathrm{d}}}\vec n} }{4\pi}\sum_{M=-1}^{1}\ 
p(S=1,\vec S\cdot\vec n=M)\;K_{s=1}(M-\vec S\cdot \vec n)\nonumber\\
&\oplus& p(S=0).
\label{formfond2part}
\end{eqnarray}
According to (\ref{formfond2part}), in order to measure $\varrho$,
we only need the probability distributions $p(S,\vec S\cdot\vec n)$,
corresponding to the operators $S^2$ and $\vec S\cdot \vec n$, for all
$\vec n$, which can be suitably recovered using the apparatus depicted 
in Fig. \ref{f:experiment2part}, which will be analyzed later.
\par
Similarly, the spin density matrix of three spins $1/2$ is
\begin{eqnarray}
\varrho=
\left( \begin{array}{cccccccc}
\xi_{11} & \xi_{12} & \xi_{13} & \xi_{14} & 0 & 0 & 0 & 0\\
\xi_{21} & \xi_{22} & \xi_{23} & \xi_{24} & 0 & 0 & 0 & 0\\
\xi_{31} & \xi_{32} & \xi_{33} & \xi_{34} & 0 & 0 & 0 & 0\\
\xi_{41} & \xi_{42} & \xi_{43} & \xi_{44} & 0 & 0 & 0 & 0\\
0 & 0 & 0 & 0 & \pi^1_{11} & \pi^1_{12} & 0 & 0 \\
0 & 0 & 0 & 0 & \pi^1_{21} & \pi^1_{22} & 0 & 0 \\
0 & 0 & 0 & 0 & 0 & 0 & \pi^2_{11} & \pi^2_{12} \\
0 & 0 & 0 & 0 & 0 & 0 & \pi^2_{21} & \pi^2_{22} \\
\end{array} \right).\label{ro3}
\end{eqnarray}
The $\xi$ block corresponds to $S=3/2$, whereas the $\pi$ blocks both
correspond to $S=1/2$, and are distinguished by their different
symmetry properties. The argument presented previously proves that
$\pi^1_{ij}=\pi^2_{ij}$, for all $i,j$, thus we can write
$\varrho=\xi\oplus\pi\oplus\pi$, with $\pi\doteq\pi^1=\pi^2$.
Again, applying (\ref{tomog}) to each block leads to
\begin{eqnarray}
\xi&=&\int\frac{{{\mathrm{d}}\vec n} }{4\pi}\sum_{M=-\frac32}^{\frac32}\ 
p(S=\frac32,\vec S\cdot\vec n=M)
\;\nonumber\\&&\times K_{s=\frac32}(M-\vec S\cdot \vec n)\,,
\\
\pi&=&\int\frac{{{\mathrm{d}}\vec n} }{4\pi}\sum_{M=-\frac12}^{\frac12} 
\frac12\;p(S=\frac12,\vec S\cdot\vec n=M)
\nonumber\\&&\times K_{s=\frac12}(M-\vec S\cdot \vec n)\,,
\end{eqnarray}
and the problem of determining $\varrho$ is again reconducted to the
simultaneous measurement of $S^2$ and $\vec S\cdot\vec n$.
\par
\begin{figure}[hbt]
\epsfxsize=.95 \hsize\leavevmode\epsffile{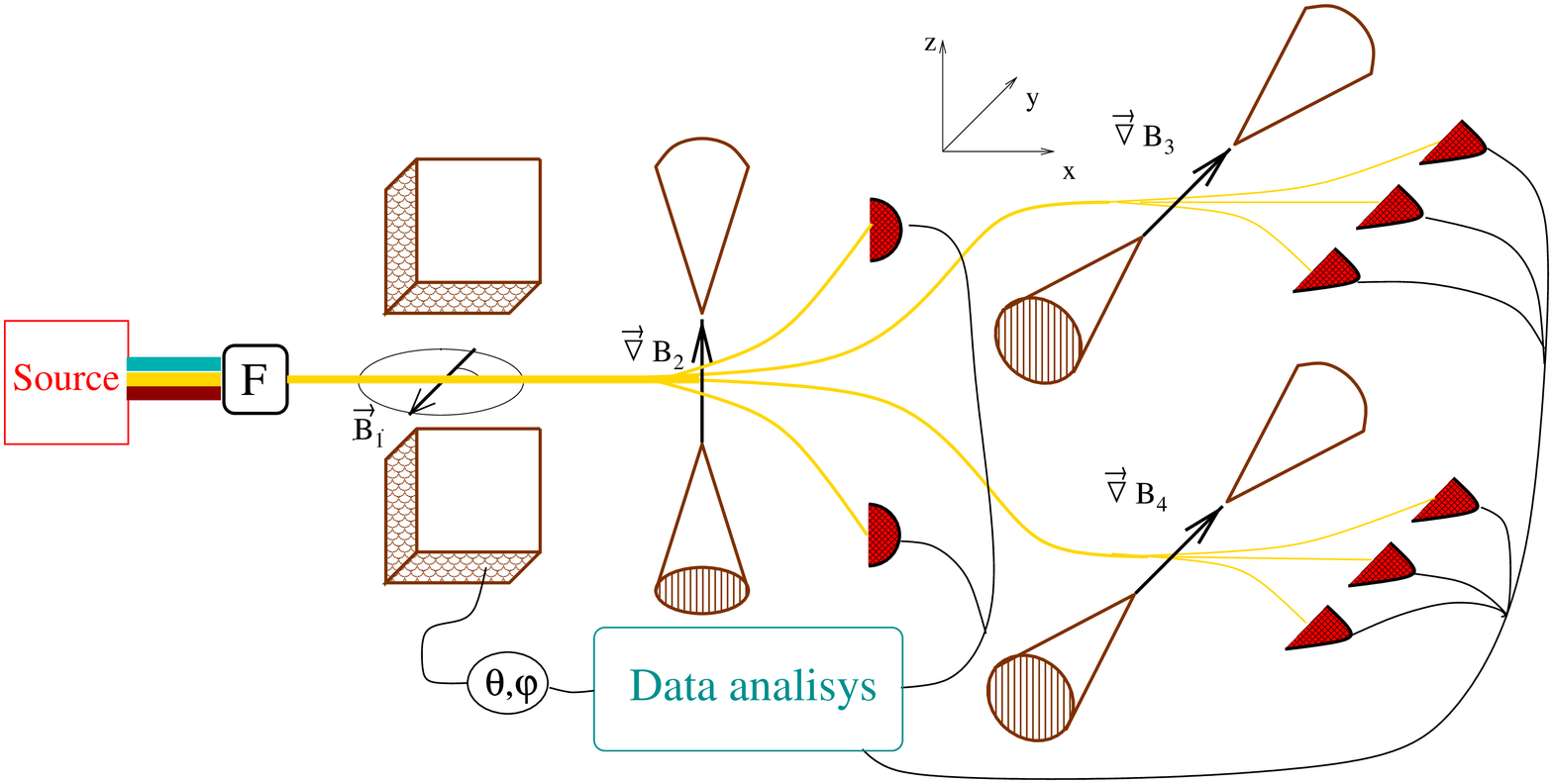}
\caption{Experimental apparatus for the tomographic reconstruction of
the spin states of systems composed of three spin $\frac 12$ states.}
\label{f:experiment3part}\end{figure}

Both in the cases presented and in the general $n$ spins case, the
required experimental data are the distributions $p(S,\vec S\cdot\vec
n)$. The apparatus to produce such data are basically equivalent in
the two cases, as evident in Figs.
\ref{f:experiment2part} and \ref{f:experiment3part}, hence we shall
limit the analysis to the two spins case. Here, the Fizeau filter and
the magnetic field $\vec B_1=B_1\vec
n_\perp=B_1(-\sin\varphi,\cos\varphi,0)$ have the same purpose as in
single particle tomography (Fig. {\ref{f:experiment}}).

Consider a beam of $n$ non--interacting systems composed of two
particles with spin $1/2$. As the analysis can be immediately extended
to a mixed case, for simplicity let us consider each system in the
pure state
\begin{eqnarray}
|\Psi_o\rangle=\gamma^s|0,0\rangle+\gamma_{-1}^a|1,-1\rangle+\gamma_{0}^a|1,0\rangle
+\gamma_{1}^a|1,1\rangle\,,\label{iniziale}
\end{eqnarray}
with $|a,b\rangle$ standing for $|S=a,M=b\rangle$.  The beam is split
into three parts by the gradient $\vec B_2$, and the systems arrive in
detector $B$ with a probability $p(S=1,M=1)=|\gamma_{1}^a|^2$ and in
detector $C$ with a probability $p(S=1,M=-1)=|\gamma_{-1}^a|^2$. The
remaining particles reach position $A$ with a probability
\begin{eqnarray}
p_A=|\gamma^s|^2+|\gamma_{0}^a|^2\label{probabA}
\end{eqnarray}
and are left in the state
\begin{eqnarray}
|\Psi_A\rangle=\frac 1{p_A}\xi\,(\gamma^s|0,0\rangle+\gamma_{0}^a|1,0\rangle).\label{AA}
\end{eqnarray}
As the subsequent gradient is directed along the $y$ axis,
Eq. (\ref{AA}) is conveniently written using the eigenstates of $S_y$,
{\it i.e.}  $|S,M\rangle_y$:
\begin{eqnarray}
&&|\Psi_A\rangle=\frac 1{p_A}\,\left[ \gamma^s
|0,0\rangle_y+\gamma_{0}^a\,\alpha_{-1}|1,-1\rangle_y
+\gamma_{0}^a\,\alpha_{0}|1,0\rangle_y\right.
\nonumber\\&&\left.+\gamma_{0}^a\,\alpha_{1}|1,1\rangle_y\right],
\end{eqnarray}
where $\alpha_i\doteq\>_y\:\langle 1,i|1,0\rangle\ (i=-1,0,1)$. Hence, the
probability for a system to arrive at detector $D$ is
\begin{eqnarray}
p_S
=\frac{1}{p_A}\,\left[|\gamma^s|^2+|{\gamma_{0}^a}|^2\,|\alpha_{0}|^2\right].
\label{probabS}
\end{eqnarray}
By measuring $p_A$ and $p_S$, the quantities $|\gamma^s|^2$ and
$|\gamma_{0}^a|^2$ are obtained by inverting equations (\ref{probabA})
and (\ref{probabS}). The coefficients $|\gamma_{0}^a|^2$,
$|\gamma^s|^2$, $|\gamma_{1}^a|^2$ and $|\gamma_{-1}^a|^2$ are the
four probabilities $p(S,M)$ we need for the reconstruction given by
Eq. (\ref{formfond2part}).

\par
A similar argument shows that the equipment of Fig.
\ref{f:experiment3part} supplies $p(S,M)$, for all $S,M$, for
a system constituted of three spins $1/2$.\par

\section{Feasibility}\label{s:feas}
The orders of magnitude of the experimental parameters are such that
the experiment is feasible with currently available technology. Only
as an example, consider the following cases of spin measurements of
electrons or nucleons. For the magnet which is responsible for the
field $\vec B_1$ with length of the order of $1$ cm, we can measure
the state of a beam of electrons with speed $\sim 10^9$ cm/sec, by
using a magnetic field $B_1=\vartheta/\gamma t$ varying between $0$
and $\sim 30$ Gauss. On the other hand, in the nucleon case, choosing
a speed of $\sim 10^7$ cm/sec, we need $B_1$ ranging between $0$ to
$\sim 10^2\div 10^3$ Gauss. Obviously, the parameters $B_1$ and $t$
can be adjusted over a wide range, according to the experimental
situation.\par

\section{Conclusions}\label{s:concl}
We have presented a tomographic experimental procedure for the
measurement of the spin density matrix. The experimental scheme is a
consequence of formula (\ref{tomog}), which was proved using group
theory. Through some Monte Carlo simulations, we have shown that the
reconstruction can be achieved with high precision using a limited
number of measurements. The extension of the procedure to the
reconstruction of states of multiple spin systems has been given, both
for distinguishable spins and for indistinguishable spin 1/2
particles. Finally, we have shown that the orders of magnitude for the
experimental setup are such that it can be implemented with currently
available technology.

\acknowledgments{This work has been sponsored by the INFM through the project
PRA-2002-CLON and by the EEC through the project IST-2000-29681
(ATESIT).}

\appendix\section{Group derivation of Quantum Tomography}\label{s:app}
The proof of Eq. (\ref{fondfor}) is obtained from the following lemma.
\par
{\it Let $A$ be an arbitrary trace--class operator on the Hilbert
space $\cal H$ of the system and $\cal R$ an irreducible unitary
square integrable representation on $\cal H$ of the tomographic group
$G$. Then }
\begin{eqnarray} 
\mbox{Tr}A=\int dg\>{\cal R}(g)A{\cal R}^\dagger(g),
\;\label{lemma1}
\end{eqnarray}
{\it where $dg$ is an invariant measure for the group $G$, normalized
as $\int dg\>|\langle u|{\cal R}(g)|v\rangle|^2=1$ which is
independent on the choice of the vectors $|u\rangle,\ |v\rangle\in
{\cal H}$.  }
\par\noindent
Proof: By using the unitarity of $\cal R$ and the properties of group
representations, we can write for any $h\in G$\begin{eqnarray}
\label{commut}
&&\int dg\>{\cal R}(g)|u\rangle\langle v|{\cal R}^\dagger(g){\cal
R}(h)
\\ \nonumber
=&&\int d{g'}\>{\cal R}(hg')|u\rangle\langle v|{\cal R}^\dagger(g')\\ 
\nonumber
=&&{\cal R}(h)
\int dg'\>{\cal R}(g')|u\rangle\langle v|{\cal R}^\dagger(g')\;,
\end{eqnarray}
which, through Schur's lemma, guarantees that \begin{eqnarray}
\int dg\>{\cal R}(g)|u\rangle\langle v|{\cal R}^\dagger(g)=\tau_{u,v}
I_{\cal H}
\;,\label{schura}
\end{eqnarray}
$I_{\cal H}$ being the identity in $\cal H$.  Consider the quantity
\begin{eqnarray}
\int dg\>|\langle u|{\cal
R}(g)|v\rangle|^2=\tau_{v,v}
\;\label{kappa},
\end{eqnarray}
where $|u\rangle$ and $|v\rangle$ are arbitrary normalized vectors in
$\cal H$.  From Eq. (\ref{schura}) it is trivial to see that
$\tau_{v,v}$ is independent on $|u\rangle$. One can check that it is
also independent on $|v\rangle$ by noticing that, given an arbitrary
vector $|a\rangle$
\begin{eqnarray}
&&\tau_{v,v}=\int dg\>{\cal R}(g)|v\rangle\langle v|{\cal
R}^\dagger(g)=\int dh\>{\cal R}(h^{-1}) |v\rangle\langle v|{\cal
R}^\dagger(h^{-1})\nonumber\\ &&=\int dh\>{\cal R}(h) |a\rangle\langle
a|{\cal R}^\dagger(h)=\tau_{a,a}
\;\label{indipkappa},
\end{eqnarray}
where the group unimodularity has been used in $dg=dh$ with $h\doteq
g^{-1}$. Notice that the hypothesis of square--integrability of the
representation guarantees the convergence of the integral in
(\ref{kappa}). Thus, the natural choice for the normalization of the
group's measure is to take $\tau_{v,v}=1$. The constant $\tau_{u,v}$
can be expressed in terms of $\tau_{v,v}$ by noticing that upon taking
$h\doteq g^{-1}$ one has
\begin{eqnarray} &&1=\tau_{v,v}=
\int dg\>\langle b|{\cal R}(g)|a\rangle\langle
a|{\cal R}^\dagger(g)|b\rangle=\nonumber\\&&\int dh\>\frac{\langle
a|{\cal R}(h)|u\rangle\langle v|{\cal R}^\dagger(h)|a\rangle}{\langle
v|u\rangle}=\frac{\tau_{u,v}}{\langle v|u\rangle}
\;\label{tauab}.
\end{eqnarray}
The lemma's thesis is now easily
found by using the Schmidt decomposition of $A$ as
$A=\sum_i\alpha_i|u_i\rangle\langle v_i|$:
\begin{eqnarray} &&\int dg\>{\cal R}(g)A{\cal
R}^\dagger(g)=\sum_i\alpha_i\int dg\>{\cal R}(g)|u_i\rangle\langle
v_i|{\cal R}^\dagger(g)=\nonumber\\
&&\sum_i\alpha_i\langle v_i|u_i\rangle=\mbox{Tr}A\;\label{tesilemma}.
\end{eqnarray}
\par

Group Tomography Theorem. {\it Let $A$ be an arbitrary trace--class
operator on the Hilbert space $\cal H$ of the system and $\cal R$ an
irreducible unitary square integrable representation on $\cal H$ of
the tomographic group $G$. Then }\begin{eqnarray} A=\int
dg\>\mbox{Tr}[A{\cal R}(g)]{\cal R}^\dagger(g)
\;\label{teortomog}.
\end{eqnarray}
\par\noindent Proof: 
Let $O$ be an invertible trace-class operator, it follows that ${\cal
R}(g)O$ is trace-class for any $g\in G$. Hence it is possible to
obtain, by applying (\ref{lemma1}) twice \begin{eqnarray}
\int dg\>\mbox{Tr}[A{\cal R}(g)]O{\cal R}^\dagger(g)=
\int dg'\>\mbox{Tr}[{\cal R}^\dagger(g')O]{\cal R}(g')A
\;\label{pa1}.
\end{eqnarray}
Take a basis $\{|k\rangle\}$ in $\cal H$, one can obtain, using again
lemma (\ref{lemma1}),
\begin{eqnarray}
&&\int dg\>\mbox{Tr}[{\cal R}^\dagger(g)O]\langle i|{\cal
R}(g)A|j\rangle=\nonumber\\
&&\int dg\>\sum_k\langle k|{\cal R}^\dagger(g)O|k\rangle\langle i|{\cal
R}(g)A|j\rangle=
\nonumber\\&&\sum_k\langle k|O\mbox{Tr}\left[|k\rangle\langle
i|\right] A|j\rangle=\langle i|OA|j\rangle 
\;\label{pa2}.
\end{eqnarray}
From Eqs. (\ref{pa1}) and (\ref{pa2}) it follows immediately that
$\int dg\>\mbox{Tr}[A{\cal R}(g)]O{\cal R}^\dagger(g)=OA$, which
yields the thesis (\ref{teortomog}) by multiplying to the left both
members by $O^{-1}$.\par
\vskip 1\baselineskip

It is trivial to extend theorem (\ref{teortomog}) to the case of
projective representations, {\it i.e.} group representations for
which, given $g_1$, $g_2$, $g_3\in G$ such that $g_1\cdot g_2=g_3$,
one has \begin{eqnarray} {\cal R}(g_1) {\cal
R}(g_2)=e^{i\zeta(g_1,g_2)}{\cal R}(g_3)
\;\label{rappraggi},
\end{eqnarray}
$\zeta\in\mathbb{R}$ being a phase factor depending on $g_1$ and
$g_2$.  Notice, moreover, that the theorem here presented is valid
also for discrete and finite groups, with the sum on group elements
replacing the integral.  From result (\ref{teortomog}), with an
appropriate choice for the tomographic group and the irreducible
representation, it is possible to prove the formula for spin
tomography (\ref{tomog}) --derived in the following section-- and for
optical homodyne tomography (\ref{homtomog}). Notice that the
unimodularity hypothesis given in the definition of tomographic group
$G$ can be relaxed without losing most of the results we give in this
paper.


\begin{thebibliography}{99}
\bibitem{haar} Haar's measure $d\mu(g)$ must be such that $\int
d\mu(g)\>=1$, and $\forall\;h\in {\cal G}$, $d\mu(g)\>=d\mu(h\cdot g)=d
\mu(g\cdot h)$.
\end{thebibliography}
\end{document}